\documentclass[journal]{IEEEtran}
\IEEEoverridecommandlockouts

\usepackage{cite}
\usepackage{amsmath}
\usepackage{bm}
\usepackage{tabularx}  
\usepackage{subcaption}
\usepackage{amssymb,amsmath}
\usepackage[mathscr]{euscript}
\usepackage{graphicx}
\usepackage{textcomp}
\usepackage{xcolor}
\usepackage{multirow}
\usepackage{booktabs}
\usepackage{algpseudocode}
\usepackage{float}
\usepackage{amsfonts}  
\usepackage[linesnumbered,ruled,vlined]{algorithm2e}

\def\BibTeX{{\rm B\kern-.05em{\sc i\kern-.025em b}\kern-.08em
    T\kern-.1667em\lower.7ex\hbox{E}\kern-.125emX}}
\begin{document}
\bstctlcite{IEEEexample:BSTcontrol}
\title{RL-Driven Security-Aware Resource Allocation for UAV-Assisted O-RAN in SAR Operations}

\author{\IEEEauthorblockN{Zaineh~Abughazzah${}^{1}$, Emna~Baccour${}^2$, 
Loay~Ismail${}^1$,
Amr~Mohamed${}^1$ and Mounir~Hamdi${}^2$} \\
\IEEEauthorblockA{
${}^1$ College of Engineering, Qatar University, Doha, Qatar.\\
${}^2$ College of Science and Engineering, Hamad Bin Khalifa University, Qatar Foundation, Doha, Qatar.\\
}
}
\maketitle

\begin{abstract}
The integration of Unmanned Aerial Vehicles (UAVs) into Open Radio Access Networks (O-RAN) enhances communication in disaster management and Search and Rescue (SAR) operations by ensuring connectivity when infrastructure fails. However, SAR scenarios demand stringent security and low-latency communication, as delays or breaches can compromise mission success. While UAVs serve as mobile relays, they introduce challenges in energy consumption and resource management, necessitating intelligent allocation strategies. Existing UAV-assisted O-RAN approaches often overlook the joint optimization of security, latency, and energy efficiency in dynamic environments. This paper proposes a novel Reinforcement Learning (RL)-based framework for dynamic resource allocation in UAV relays, explicitly addressing these trade-offs. Our approach formulates an optimization problem that integrates security-aware resource allocation, latency minimization, and energy efficiency, which is solved using RL. Unlike heuristic or static methods, our framework adapts in real-time to network dynamics, ensuring robust communication. Simulations demonstrate superior performance compared to heuristic baselines, achieving enhanced security and energy efficiency while maintaining ultra-low latency in SAR scenarios.

\end{abstract}


\section{Introduction} \label{sec:introduction}
UAVs are increasingly being considered as key enablers for enhancing communication networks in critical scenarios, such as SAR operations \cite{uav_sar_survey}. In such high-risk scenarios, reliable communication is crucial for coordinating rescue efforts, yet damaged base stations and limited connectivity often hinder real-time information exchange. UAVs offer a rapid and adaptable solution, providing dynamic network coverage where conventional systems fail. Their seamless integration into the O-RAN architecture, which supports open interfaces and cloud-based flexibility, enables efficient and scalable SAR operations \cite{9839628}. However, several challenges must be addressed to fully harness their potential. Energy efficiency is a primary concern, as UAVs operate on limited battery life, requiring optimized resource allocation to maximize flight duration and maintain continuous coverage. Security is another critical factor—any data breach or unauthorized access could jeopardize rescue missions, making robust encryption and authentication mechanisms essential. Furthermore, low-latency communication is imperative for real-time decision-making, ensuring that mission-critical data reaches responders without delays. In disaster-prone environments, mitigating noise and maintaining a low BER are also crucial for reliable communication. By tackling these challenges, UAV-assisted O-RAN can significantly enhance network resilience, ensuring secure, efficient, and uninterrupted communication in SAR operations.
Research on UAV-based relays in O-RAN for SAR operations has gained momentum due to the increasing demand for resilient communication in disaster recovery. Previous studies have focused on UAV integration in wireless networks, addressing coverage extension, latency minimization, and energy efficiency as well as resource allocation. In \cite{10584067}, an O-RAN-enabled UAV-assisted network optimizes radio unit association, deployment, and resource allocation using a double-loop algorithm that combines the Dinkelbach method and Block Coordinate Descent. Similarly, \cite{9522072} explores UAV relay-assisted emergency communications in IoT networks, optimizing bandwidth, power allocation, and UAV trajectory while considering latency constraints and self-interference mitigation. In \cite{9712640}, physical-layer security in UAV-assisted IoT networks is enhanced by optimizing UAV trajectory, scheduling, and power allocation to maximize secrecy rates. 
In \cite{10107729}, a multi-agent RL-based UAV swarm communication scheme optimizes relay selection and power allocation to combat jamming, improving efficiency and robustness through policy and transfer learning.
To the best of our knowledge, existing research on UAV-based relays in O-RAN for SAR operations overlooks the trade-off between encryption-based security and latency, especially in noise suppression for secure communication. While UAVs improve coverage, ensuring secure, low-latency communication remains a challenge. This calls for a Multi-Objective Optimization Problem (MOP) to balance conflicting objectives—security, latency, and energy efficiency—since optimizing  a single objective (e.g., latency) might compromise others. Our framework fills this gap by using RL to dynamically allocate resources, optimizing security, power efficiency, and latency in real-time, ensuring an adaptive, secure, and efficient communication network for SAR operations.

The main contributions of this paper are as follows:
\begin{itemize}
\item We propose a dynamic resource allocation framework for O-RAN, optimizing the trade-off between security, power efficiency, and communication latency in UAV-based relays for SAR operations.
\item We formulate a MOP for association and security level selection, aiming to minimize total latency and energy consumption while maximizing system security.
\item We introduce an RL-driven approach to solve the MOP, leveraging sub-optimal solutions for efficient real-time resource allocation while dynamically adjusting encryption levels to balance power consumption, latency, and security. Specifically, we employ Proximal Policy Optimization (PPO) for multi-objective optimization.
\item We conduct performance analysis through simulations to demonstrate the effectiveness of the RL-based framework.
\end{itemize}

\section{System model} \label{sec:system_model}

\begin{figure}[h]
    \centering
    \includegraphics[width=1\linewidth]{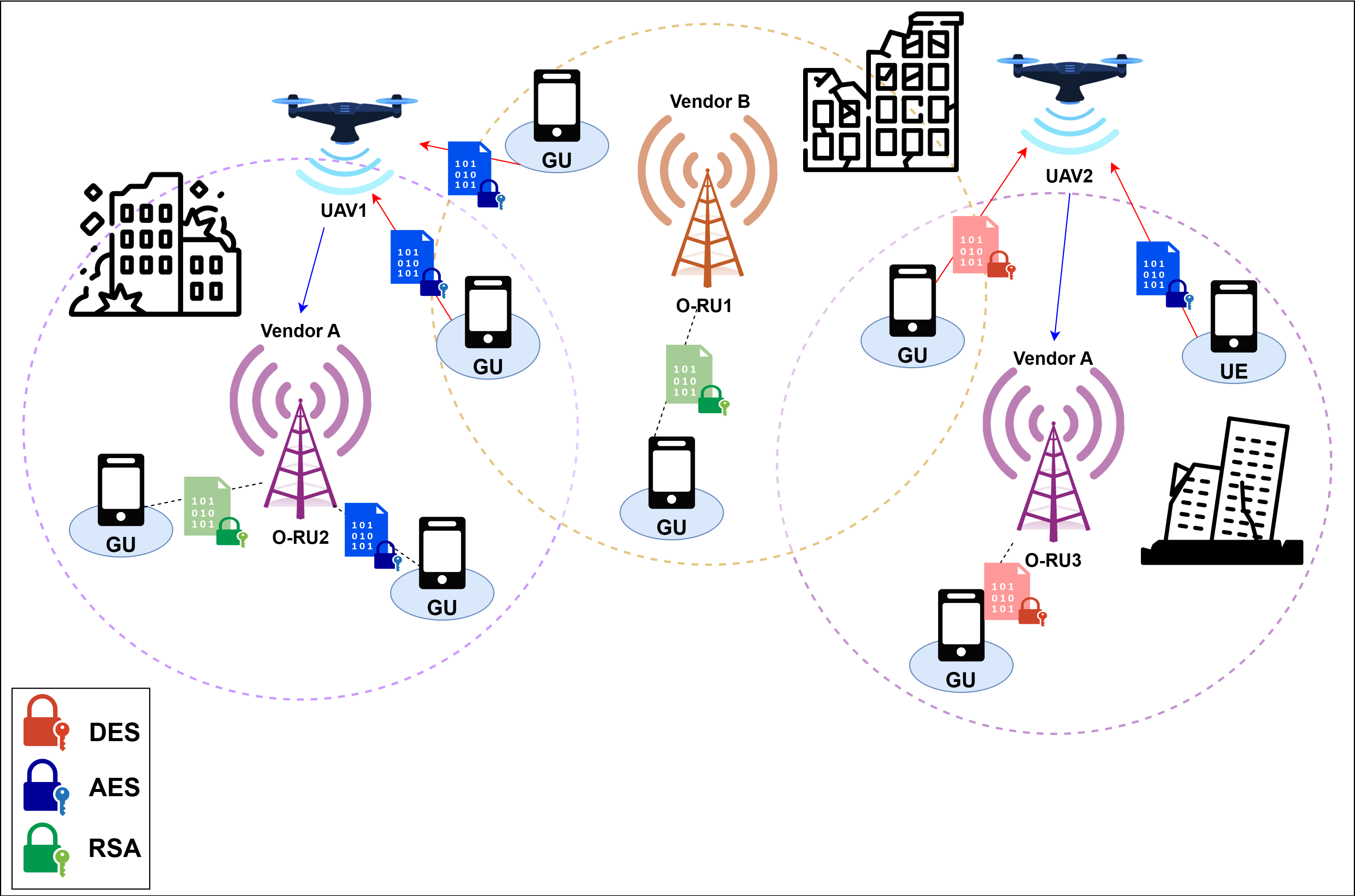}
    \caption{System Model}
    \label{fig:system_model}
\end{figure}

As shown in Figure \ref{fig:system_model}, we consider an uplink wireless 5G O-RAN network consisting of a set of UAV-based relays denoted as $\mathcal{A} = \{1, 2, ..., A\}$, where $A$ represents the total number of UAVs deployed in the SAR scenario to ensure communication coverage in the affected area. Each UAV relay can connect to an O-RAN Radio Units (O-RU) and equipped with a transmission power denoted as \(P_a\), which defines its capability to relay data between Ground Users (GUs) and O-RUs, without requiring significant computational power.  
Moreover, each UAV has a trajectory denoted as $q_a^t$, which represents the position of the UAV at each timestep. It affects the connectivity and latency of communication between the UAVs, GUs, and O-RUs, influencing the energy efficiency, security, and resource allocation strategies.

Additionally, we assume the presence of a set of O-RUs denoted as $\mathcal{G} = \{1, 2, ..., G\}$, where $G$ is the total number of O-RUs. Let $\mathcal{P}_{g}$ denote the processor clock speed of the O-RU. Each O-RU has its own computational and security requirements, denoted as $W_g$ for the security level, and Resource Blocks (RBs) denoted as $M_g$, representing the resources available for data transmission. Moreover, we define a set of GUs, represented as $\mathcal{u} = \{1, 2, ..., U\}$, where $U$ is the total number of GUs involved in the SAR operation. Let $Q_u$ represent the processor’s clock speed GU u, and $Z_u$ denote battery capacity, while $\Gamma_u$ indicates its computational capabilities. Each GU $u \in \mathcal{U}$ is connected to one O-RU $g \in \mathcal{G}$ at each time step $t \in \mathcal{T}$ directly through a UAV, where $\mathcal{T}$ represents the total time steps throughout the operation.
At each time step $t$, each GU $u$ aims to send data of size $D_{u,t}$ in the uplink direction. The selection of the O-RU for each GU depends on factors such as security requirements, available bandwidth, user mobility, and the trajectory of the UAVs, evaluated at each time step. Similarly, UAVs select the O-RU based on these conditions to enhance communication. UAVs maintain continuous connectivity to O-RUs using dedicated resources, ensuring uninterrupted communication. Given the battery constraints of both UAVs and GUs, energy-efficient communication and resource management, including the optimization of UAV trajectories, are crucial to balance security, power consumption, and latency, ensuring reliable and secure connectivity during SAR operations.

\subsection{Encryption Model} \label{Encryption_algorithms}

For secure data transmission between GUs and O-RUs, we adopt encryption algorithms to ensure confidentiality. Building on the approach in \cite{Abughazzah2025}, we assume that key exchange protocols are established and focus on three widely used encryption schemes: Data Encryption Standard (DES), Advanced Encryption Standard (AES), and Rivest-Shamir-Adleman (RSA). Each algorithm has a specific key length set, influencing security and computational efficiency. The plaintext $PT_{u,t}$ is divided into fixed-size blocks and encrypted using the Electronic Code Book (ECB) mode for simplicity.
The encrypted ciphertext $CT_{u,t}$ is transmitted to the designated O-RU, where it is decrypted using the corresponding algorithm and key. 
For clarity, we define the variables as follows:
The list of encryption algorithms is denoted by 
$ k \in \mathcal{K} = \{1,2,3\}$, where:
\begin{footnotesize}
\begin{equation}\label{eq:algorithms}
    k = \left\{\begin{matrix}
\text{$1$, DES algorithm}\\
\text{$2$, AES algorithm}\\
\text{$3$, RSA algorithm}
\end{matrix}\right. 
\end{equation}
\end{footnotesize}
The key lengths for each algorithm are denoted by
$D_k$, where:  
\begin{footnotesize}
\begin{equation}\label{eq:key_lengths}
    D_k = \left\{\begin{matrix}
64 \text{ bits} & \text{for DES} \\
      128, 192, 256 \text{ bits} & \text{for AES} \\
      1024, 2048, 3072, 4096 \text{ bits} & \text{for RSA} \\
\end{matrix}\right. 
\end{equation}
\end{footnotesize}
The block size for each algorithm \( k \) is given by \( \mathcal{B}_k \), where:
\begin{footnotesize}
\begin{equation}\label{eq:algorithms}
   \mathcal{B}_k  = \left\{\begin{matrix}
\text{$\mathcal{B}_{DES}$ = 64}\\
\text{$\mathcal{B}_{AES}$ = 128}\\
\text{$\mathcal{B}_{RSA}$ = $N$}
\end{matrix}\right. 
\end{equation}
\end{footnotesize}

\subsection{Computational Latency Model} \label{Computation_model}

In our previous work \cite{Abughazzah2025}, we developed a computational model to analyze the encryption and decryption latencies and energy consumption based on the complexities of different encryption algorithms (DES, AES, and RSA). The computational complexities $C_k$ for each algorithm is given by:
\begin{footnotesize}
\begin{equation}
    C_{k} = \left\{
    \begin{matrix}
    C_{DES_{enc}} = C_{DES_{dec}}= 16 \cdot N_{shift}\\ + R_{DES}( 10 \cdot N_{shift} +  10 \cdot N_{xor})\\
    \\
    C_{AES_{enc}} = 16 \cdot N_{xor} + (R_{AES}-1)( 184 \cdot N_{and}\\ +  136 \cdot N_{or} + 352 \cdot N_{shift})\\ + ( 16 \cdot N_{xor} +  12 \cdot N_{shift} + 12 \cdot N_{or})\\
    \\
    C_{AES_{dec}} = 16 \cdot N_{xor} + (R_{AES}-1)( 644 \cdot N_{and} \\+  500 \cdot N_{or} + 224 \cdot N_{shift}) \\+ ( 16 \cdot N_{xor} +  12 \cdot N_{shift} + 12 \cdot N_{or})\\
    \\
     C_{RSA_{enc}} = C_{RSA_{dec}} = N^2\\
\end{matrix}\right.
\end{equation}
\end{footnotesize}
where $N$ is the key length, and $R_k$ is the number of rounds for each algorithm  (i.e., 16 in DES, 10 in AES-128, 12 in AES-192, and 14 in AES-256) while \( N_{\text{and}}, N_{\text{or}}, N_{\text{shift}}, N_{\text{xor}} \) represent the number of cycles required to execute a single AND, OR, shift, and XOR operation, respectively.
The latencies for encryption done by the user and decryption processed by the O-RU are defined as:
\begin{footnotesize}
\begin{equation}
\tau_{enc}^{u,k,t} =  C_k \frac{\lceil \frac{D_{u,t}}{\mathcal{B}_{k}} \rceil}{\mathcal{Q}_{u}}, ~\footnotesize
\tau_{dec}^{g,k,t} = C_k \frac{\lceil \frac{D_{u,t}}{\mathcal{B}_{k}} \rceil}{\mathcal{P}_{g}}
\end{equation}
\end{footnotesize}
\subsection{Communication Model}  
\label{Communication_model}
The communication model includes both communication latency and BER models, which are essential for evaluating UAV-assisted communication systems and ensuring efficient, reliable, and secure data transmission.

\subsubsection{Communication Latency}
The communication latency describes the transmission delay when offloading data from the GU $u$ to the O-RU $g$, which can occur either directly or indirectly through a UAV.

The distance between devices impacts the transmission rate and latency. The Euclidean distances for direct and indirect transmission are given by:  
\vspace{-0.5em}
\begin{footnotesize}
\begin{equation}
    d_{g,u} = \sqrt{H_g^2 + \| \omega_g - \omega_u \|^2}
\end{equation}
\end{footnotesize}
\vspace{-0.5em}
\begin{footnotesize}
\begin{equation}
    d_{a,u} = \sqrt{H_a^2 + \| q_a - \omega_u \|^2}
\end{equation}
\end{footnotesize}
\vspace{-0.5em}
\vspace{-0.5em}
\begin{footnotesize}
\begin{equation}
    d_{a,g} = \sqrt{\| H_g -  H_a \|^2 + \| \omega_g - q_a \|^2}
\end{equation}
\end{footnotesize}
where the O-RU and GU projections are \(\omega_g\) and \(\omega_u\), respectively, while the UAV relay is \(q_a\). \(H_a\) and \(H_g\) represent the UAV's altitude and O-RU's antenna height, respectively.

Hence, the communication latencies can be expressed as:

\begin{footnotesize}
\begin{equation}
    \tau_{u,g}^t = \frac{D_{u,t}}{B_{u,g} \log_2 \left( 1 + \frac{p_{u,g} h_{u,g,t}}{\sigma^2} \right)}
\end{equation}
\end{footnotesize}
\vspace{-0.7em}
\begin{footnotesize}
\begin{equation}
    \tau_{u,a}^t = \frac{D_{u,t}}{B_{u,a} \log_2 \left( 1 + \frac{p_{u,a} h_{u,a,t}}{\sigma^2} \right)}
\end{equation}
\end{footnotesize}
\vspace{-0.7em}
\begin{footnotesize}
\begin{equation}
    \tau_{a,g}^t = \frac{D_{u,t}}{B_{a,g} \log_2 \left( 1 + \frac{p_{a,g} h_{a,g,t}}{\sigma^2} \right)}
\end{equation}
\end{footnotesize}
where $B$ is the bandwidth, $\sigma^2$ is the noise power, $p$ is the transmission power, $h = \frac{\beta_0}{d^{\alpha}}$ is the channel gain where $\beta_0$ is the path loss at unit distance, d depends on the communicating devices, and $\alpha$ is the path loss exponent. It is worth mentioning that the Signal-to-Noise Ratio is given by $SNR = \frac{P * h}{\sigma^2}$.

\subsubsection{BER Model}

BER is a critical metric for evaluating the quality of communication between GUs and the O-RUs. A low BER indicates good communication quality, while a high BER signifies poor signal quality and unreliable data transmission. For Binary Phase Shift Keying (BPSK), the BER as a function of Signal-to-Noise Ratio (SNR) for a GU \( u \) at time step \( t \) is expressed as: $ \text{BER}_{\text{BPSK}} = \frac{1}{2} \cdot \text{erfc}\left(\sqrt{\text{SNR}}\right)$
where higher SNR reduces the BER, improving communication reliability. UAVs optimize their position to enhance SNR, minimize interference, and reduce BER. Additionally, a high BER increases the risk of transmission errors, potentially compromising data integrity and system security, especially in encrypted communication.

\subsection{Security model}
\label{Security_model}
The security model defines data security during transmission from the GU to the O-RU, depending on the encryption technique and key size. Larger keys enhance security by making cryptanalysis harder but also increase computational overhead, potentially slowing communication and consuming more resources—especially on constrained devices.
The encryption robustness is expressed as \( S(N) = \log_2(N) \), where \( N \) is the key size. Thus, the security level achieved by GU \( u \in \mathcal{U} \) at time \( t \in \mathcal{T} \) is:
\begin{footnotesize}
\begin{equation}
    S_{u,t}= \log_2(N_{u}^{t}), \forall u \in \mathcal{U}, \forall t \in \mathcal{T}
\end{equation}
\end{footnotesize}

where \( N_{u,t} \) is the key length used by GU \( u \) at time \( t \).



\subsection{Mechanical Energy model}

In this work, we focus on the mechanical energy consumption of rotary-wing UAVs, which dominates over processing energy. Unlike fixed-wing UAVs, they require constant power for hovering and maneuvering, making energy use highly dependent on velocity and flight dynamics. The total mechanical energy consumption over a time period is given by \cite{uav_energy}:
\begin{footnotesize}
\begin{equation}
\begin{aligned}
    E_a &= \sum_{t=1}^{T} \delta_t \left( P_0 \left( 1 + \frac{3 v_{a,t}^2}{U_{\text{tip}}^2} \right) + c_0 v_{a,t}^3 \right) \\
    &\quad + \sum_{t=1}^{T} \delta_t P_1 \left( \sqrt{1 + \frac{v_{a,t}^4}{4 v_0^4}} - \frac{v_{a,t}}{2 v_0^2} \right)
\end{aligned}
\end{equation}
\end{footnotesize}
where \( P_0 \) and \( P_1 \)re the Blade Profile Power and Induced Power of all UAVs in hover, \( c_0 \) is a constant for Parasite Power, \( \delta_t \) is the time duration at time slot \( t \), \( U_{\text{tip}} \) is the rotor blade tip speed,  \( v_0 \) is the motor-induced velocity, \( v_{a,t} \)  is the UAV's instantaneous velocity at time step \( t \), given by:
\begin{footnotesize}
\begin{equation}
    v_a^t = \frac{\sqrt{(d_{a,x}^t)^2 + (d_{a,y}^t)^2 + (d_{a,z}^t)^2}}{\delta_t}
\end{equation}
\end{footnotesize}
where \( d_{a,x}^t \), \( d_{a,y}^t \), and \( d_{a,z}^t \) represent the displacement of UAV \(a\) at time step \( t \).


\section{Problem Formulation} \label{sec:problem_formulation}
This work aims to associate each GU $u \in \mathcal{U}$ with an O-RU $g \in \mathcal{G}$ at each time step $t \in \mathcal{T}$, either directly or via a UAV $a \in \mathcal{A}$, minimizing latency and energy consumption while maximizing security. This involves a trade-off: higher security levels increase latency due to encryption overhead, while energy constraints limit UAV operation. Efficient resource management is critical due to varying O-RU security requirements, as well as the computational power and battery life limitations of GUs. Balancing security, energy efficiency, and latency is essential for seamless SAR operation connectivity.


To formulate this joint problem, we define the MOP with four decision variables. We model the association binary decision variables by $x_{u,g}^{t}$ and $y_{u,a,g}^{t}$ defined as:
\begin{footnotesize}
\begin{equation}
    x_{u,g}^{t} = \left\{\begin{matrix}
\text{1, if GU $u \in \mathcal{U}$ is associated directly with}\\ \text{O-RU $g \in \mathcal{G}$ at time step $t \in \mathcal{T}$}\\
\text{0, otherwise}
\end{matrix}\right.
\end{equation}
\end{footnotesize}
\begin{footnotesize}
\begin{equation}
    y_{u,a,g}^{t} = \left\{\begin{matrix}
\text{1, if GU $u \in \mathcal{U}$ is associated with O-RU $g \in \mathcal{G}$ indirectly} \\ \text{through UAV $a \in \mathcal{A}$ at time step $t \in \mathcal{T}$}\\
\text{0, otherwise}
\end{matrix}\right.
\end{equation}
\end{footnotesize}

Moreover, we model the selection of the key length applied by GU $u \in \mathcal{U}$ at time step $t \in \mathcal{T}$ with the variable $N_{u}^{t}$. The key length must belong to the union of the possible options for all encryption algorithms ${\mathcal{D}_k}$:
\begin{footnotesize}
\begin{equation}
{{N}_{u}^{t}} \in \cup_{k=1}^{K}{\mathcal{D}_k}, \quad \forall u \in \mathcal{U}, \forall t \in \mathcal{T}
\end{equation}
\end{footnotesize}
We omitted the algorithm selection since there is no overlapping between key lengths in the encryption algorithms we adopt for this study. Hence, by choosing the key length, the algorithm can be derived.


Lastly, to incorporate UAV mobility, we define trajectory decision variable $q_{a}^{t}$ as the position of UAV $a \in \mathcal{A}$ at time step $t \in \mathcal{T}$. Trajectory decisions impact latency, energy consumption, and security by influencing movement cost and channel quality. 

From the discussion in Section \ref{sec:system_model}, the total latency to transmit data from GU $u$ to the selected O-RU $g$ at a time step $t$ is given by:
\vspace{-0.5em}
\begin{footnotesize}
\begin{equation}
    L_{u,t}= \tau_{enc}^{u,t} + \tau_{comm}^{u,t} + \sum_{g=1}^{G}{x_{u,g}^{t}} \cdot \tau_{dec}^{g,t}
\end{equation}
\end{footnotesize}
where communication latency is given by:
\begin{footnotesize}
\begin{equation}
    \tau_{\text{comm}}^{u,t} = \sum_{g=1}^{G} \sum_{a=1}^{A} {x_{u,g}^{t} \cdot \tau_{u,g}^t + y_{u,a,g}^t \cdot (\tau_{a,u}^t + \tau_{a,g}^t)}
\end{equation}
\end{footnotesize}

The problem formulation, aiming at minimizing latency and UAV energy consumption while maximizing security through optimized GU association, encryption key lengths, and UAV trajectories, is presented in Equation (\ref{eq:P1}). Note that all objectives are weighted and normalized between 0 and 1 by dividing by their maximum values to ensure a fair trade-off between the objectives. The notation $[\cdot]^{\sim}$ represents the normalization and weighting function.


\begin{subequations}\label{eq:P1}
\footnotesize
\allowdisplaybreaks
\begin{align} 
\textbf{P1: } &\min_{\bm{x_{u,g}^{t},y_{u,a,g}^{t},N_{u}^{t},q_{a}^{t}}} \nonumber \\ & \sum_{t=1}^{T}{\sum_{a=1}^{A}{[E_{a,t}]^{\sim}}} + \sum_{t=1}^{T}{\sum_{u=1}^{U}{[L_{u,a,g,t}]^{\sim}}} + \sum_{t=1}^{T}{\sum_{u=1}^{U}{1-[S_{u,t}]^{\sim}}}\\
 \text{s.t.} &\sum_{g=1}^{G}( {x_{u,g}^{t}} + \sum_{a=1}^{A}{y_{u,a,g}^{t}}) = 1, \forall u \in \mathcal{U} \enspace \forall t \in \mathcal{T} \label{eq:association3} \\
 & \sum_{g=1}^{G} W_{g} \cdot ( {x_{u,g}^{t}} + \sum_{a=1}^{A}{y_{u,a,g}^{t}}) \leq S_{u,t} , \forall u \in \mathcal{U} \enspace \forall t \in \mathcal{T} \label{eq:Smin}\\
 &\sum_{g=1}^{G} \sum_{u=1}^{U}{y_{u,a,g}^{t}} \leq M_{a} , \forall a \in \mathcal{A} \enspace \forall t \in \mathcal{T} \label{eq:resource_blocks1} \\
 &\sum_{u=1}^{U} {x_{u,g}^{t}} \leq M_{g} , \forall g \in \mathcal{G} \enspace \forall t \in \mathcal{T} \label{eq:resource_blocks2} \\
   &  \sum_{u=1}^{U} { N_{u}^{t} \cdot \mathcal{C}_{u,t}}  \leq \Gamma_u, \forall t \in \mathcal{T} \label{eq:computational_power} \\
  &\sum_{t=1}^{T}{[ (\eta_{cp}^{u,t}) + (\eta_{cm}^{u,t})]} \leq Z_u, \forall u \in \mathcal{U} \label{eq:power_consumption} \\
   & \gamma_{u}^{t} \leq \text{BER}_{\text{max}}, \quad \forall u \in \mathcal{U}, \forall t \in \mathcal{T} \label{BER} \\
 & \|\mathbf{q}_i^{t} - \mathbf{q}_j^{t}\| \geq d_{\text{min}}, \quad \forall i, j \in \mathcal{A}, \, i \neq j, \, \forall t \in \mathcal{T} \label{collision_constraint} \\
 & d_{a,t,t+1} \leq d_{\text{max}}, \quad \forall a \in \mathcal{A}, \forall t \in \mathcal{T} \label{max_dist}\\
 &{{N}_{u}^{t}} \in \cup_{k=1}^{K}{\mathcal{D}_k}, \quad \forall u \in \mathcal{U}, \forall t \in \mathcal{T} \label{eq:n_value2} \\
 &x_{u,g}^{t}, y_{u,a,g}^{t} \in \{0,1\}, \forall u \in \mathcal{U}, \forall g \in \mathcal{G}, 
 \forall a \in \mathcal{A}, \forall t \in \mathcal{T} \label{A_x_lim}
\end{align}
\end{subequations}
where $\eta_{cp}^{u,t}$ and $\eta_{cm}^{u,t}$ represent the power needed for the computations and communication respectively: 
\begin{footnotesize}
\begin{equation}
    \eta_{cp}^{u,t}= \tau_{enc}^{u,k,t} \cdot E_{cp}, \quad \eta_{cm}^{u,t}= \tau_{comm}^{u,a,g,t} \cdot E_{cm}
\end{equation}
\end{footnotesize}
where $E_{cp}$ and $E_{cm}$ represent the computation power consumption and transmission power consumption per second.

Constraint (\ref{eq:association3}) ensures each GU is connected to exactly one O-RU, either directly or via a UAV. Constraint (\ref{eq:Smin}) guarantees each GU's security level meets the O-RU’s requirement. Constraints (\ref{eq:resource_blocks1}) and (\ref{eq:resource_blocks2}) limit the number of GUs connected to O-RUs and UAVs to their available RBs. Constraints (\ref{eq:computational_power}) and (\ref{eq:power_consumption}) ensure that computational and battery resource usage on GUs stays within limits. Constraint (\ref{eq:n_value2}) enforces valid key lengths from \( \mathcal{Dk} \). Constraint (\ref{A_x_lim}) ensures \( x_{u,g}^{t} \) and \( y_{u,a,g}^{t} \) are binary. Constraint (\ref{BER}) limits the BER \( \gamma_{u}^{t} \) for GU \( u \) at time \( t \). Constraint (\ref{collision_constraint}) prevents UAV collisions by maintaining a minimum distance. Lastly, constraint (\ref{max_dist}) limits each UAV’s displacement per time step.

\section{Solution} \label{sec:solution}
To address UAV trajectory management, resource allocation, and device associations in dynamic wireless networks, we propose a model-free Deep Reinforcement Learning (DRL) approach. Due to the NP-hard nature of the problem—caused by unknown demands, mobility, and varying conditions, traditional optimization methods are impractical. We adopt PPO, an on-policy RL technique suited for complex problems.  PPO ensures stable convergence by using a clipped objective to prevent large, costly updates, and it adapts in real time without prior system knowledge \cite{continuousDRL}.

\subsection{MDP formulation}
To manage UAV-assisted wireless communication and resource allocation, we model the problem as a Markov Decision Process (MDP) with state and action spaces and a reward function, aligning with the continuous mixed-integer decision space of the PPO algorithm. The training spans \( N \) episodes, each representing a time period \( T \) during which users exchange data with O-RUs. Episodes are independent, resetting cumulative rewards, allowing the agent to refine its strategy through past actions, rewards, and performance analysis.

\subsubsection{State Space ($S$)} represents the system’s status at each time step, including the location and battery levels of GUs, the data size to be transmitted, and the UAVs’ previous location coordinates, providing the RL agent with necessary information for optimal decisions.

\subsubsection{Action Space ($A$)} involves selecting actions that determine UAV displacement, GU association with O-RUs directly or via UAV relays, and GU key length to optimize communication efficiency while adhering to system constraints.

\subsubsection{Reward Function ($R$)} evaluates performance, considering energy consumption, latency, and security, aligning with optimization goals. It encourages a balance between these factors, with penalties for constraint violations, guiding the agent toward feasible solutions. The positive reward \( r_t \) at each iteration \( t \) is defined as:
\begin{footnotesize}
\begin{equation}
r = w_1 \cdot (1-\text{Latency}) + w_2 \cdot (1-\text{Energy}) + w_3 \cdot \text{Security}
\end{equation}
\end{footnotesize}
where Latency, Energy, and Security are the normalized objectives, and \( w_1, w_2, w_3 \) are the weight factors for each objective. The penalty term \( p_t \) accounts for constraints' violations at time step t. The penalty function is defined as:
\begin{footnotesize}
\begin{equation}
\label{eq:penalty}
p_t = p_{\text{security}_t} + p_{\text{RBs}_t} + p_{\text{comp}_t} + p_{\text{battery}_t} + p_{\text{BER}_t} + p_{\text{collision}_t} + p_{d_{\text{max}_t}}
\end{equation}
\end{footnotesize}
where each term penalizes a constraint violation: \( p_{\text{security}_t} \) ensures security levels meet requirements, \( p_{\text{RBs}_t} \) addresses resource block allocation violations, \( p_{\text{comp}_t} \) and \( p_{\text{battery}_t} \) prevent excessive computational and battery consumption by GUs, \( p_{\text{BER}_t} \) penalizes high BER, which can lead to poor communication, \( p_{\text{collision}_t} \) ensures UAVs maintain safe distances to avoid collisions, and \( p_{d_{\text{max}_t}} \) limits UAV displacement.


The final reward, accounting for penalties, is defined as:
\begin{footnotesize}
\begin{equation}
\label{eq:reward_final}
r_{\text{final},t} = r_t - p_t
\end{equation}
\end{footnotesize}




\subsubsection{Agent Design}
The agent learns the optimal policy \( \pi : S \times A \to [0, 1] \) to maximize total rewards over episodes by approximating the action-value function \( Q_{\pi}(s, a) \):
\begin{footnotesize}
\begin{equation}
    Q_{\pi}(s, a) = \mathbb{E}\left[\sum_{k=1}^{\infty} \gamma^k r_{t+k+1} \mid s_t = s, a_t = a \right]
    \label{eq:qfunc}
\end{equation}
\end{footnotesize}
where \( \gamma \in [0, 1] \) is the discount factor that balances the importance of immediate and future reward. The optimal policy maximizes the Q-value is defined as:
\begin{footnotesize}
\begin{equation}
    \pi^* = \arg \max_a Q_{\pi}(s, a)
\end{equation}
\end{footnotesize}
\subsection{RL-Based Optimization Using PPO}
In order to improve the exploration capability of policy-based methods, model-free and on-policy learning is employed. In this approach, learning relies on past actions and the current policy, without requiring prior knowledge of the environment. One of the most popular on-policy algorithms introduced by OpenAI is PPO \cite{ppo_algorithms}. The objective function for PPO is given by:
\begin{footnotesize}
\begin{equation}
\label{eq:ppo_objective}
L^{CLIP}(\theta) = \mathbb{E} \left[ \left( m_t(\theta) \hat{A}_t, \text{clip}(m_t(\theta), 1 - \epsilon, 1 + \epsilon) \hat{A}_t \right) \right]
\end{equation}
\end{footnotesize}
where \( \epsilon \) is the clipping parameter that helps to prevent large policy updates, \( \mathbb{E} \) is the expected value over the trajectories sampled from the environment, \( m_t(\theta) \) is the policy probability ratio between the new and old policies and is given by:
\begin{footnotesize}
\begin{equation}
m_t(\theta) = \frac{\pi(a_t | s_t,\theta)}{\pi(a_t | s_t,\theta_p)}
\end{equation}
\end{footnotesize}
and \( \hat{A}_t \) is the estimator for the advantage function at time step \( t \) and is given by:
\vspace{-0.5em}
\begin{footnotesize}
\begin{equation}
\label{eq:Advantage_Estimates}
\hat{A}_t = \sum_{l=0}^{\infty} (\gamma \lambda)^l \delta_{t+l}
\end{equation}
\end{footnotesize}
\begin{footnotesize}
\begin{equation}
\label{eq:Advantage_est2}
\delta_{t+l} = r_t + \gamma V(s_{t+1}, \theta) - V(s_t, \theta)
\end{equation}
\end{footnotesize}
where $V(s_t, \theta)$ denotes the expected reward, calculated by averaging Eq. \ref{eq:qfunc}  across all possible actions at state $s_t$, and $\lambda$ is the learning rate that controls the bias-variance trade-off.


The pseudo-code of the proposed PPO-based algorithm is shown in Algorithm \ref{algo:PPO_algo}. The algorithm runs for \( N \) episodes, each consisting of \( T \) iterations (time steps). At each step, the RL agent makes several decisions, including associating GUs with O-RUs (directly or via a UAV), selecting the optimal encryption key length, and determining the UAV trajectory. The goal is to minimize latency and energy consumption while maximizing security. 
The PPO algorithm begins by initializing system parameters as wll as two identical policies with random weights: $\theta$, which is trained, and $\theta_p$, which is updated gradually to ensure a monotonic improvement of the policy $\pi(\theta_p)$. At each time step, the agent observes the state, selects an action, computes penalties and rewards, and stores the experience. The policy $\theta$ is then updated by maximizing a clipped surrogate objective, and changes to $\theta_p$ are limited to promote stable learning and smooth convergence.

\begin{footnotesize}
\begin{algorithm}
\caption{PPO Algorithm}
\label{algo:PPO_algo}
\textbf{Initialization:}\;
Initialize the locations for O-RUs and their security requirements, GUs capabilities (i.e., battery, computational power), Data size\;
\textbf{PPO Training:}\;
Initialize $\theta$ with a random value\;
$\theta_p \gets \theta$\;
\For{$n = 1$ \KwTo $N$}{
    \For{$t = 1$ \KwTo $T$}{
        Obtain current \textbf{State} $s_t$\;
        Select \textbf{Action} $a_t$\;
        Calculate \textbf{Penalty} from Eq.~\ref{eq:penalty} $p_t$\;
        Observe \textbf{Reward} from Eq.~\ref{eq:reward_final} $r_{\text{final},t}$\;
        Observe \textbf{Next State} $s_{t+1}$\;
        Save ($s_t$, $a_t$, $r_t$, $s_{t+1}$) in the experience memory $\mathbb{D}$\;
    }
    Evaluate $\hat{A}_t$ from Eq.~\ref{eq:Advantage_Estimates} and Eq.~\ref{eq:Advantage_est2}\;
    Sample a mini-batch of ($s_t$, $a_t$, $r_t$, $s_{t+1}$) from $\mathbb{D}$\;
    Update $\theta$ by maximizing Eq.~\ref{eq:ppo_objective}\;
    Update Previous policy: $\theta_p \gets \theta$\;
}
\end{algorithm}
\end{footnotesize}
\section{Results} \label{sec:Results}
This section evaluates our RL-based approach, which employs a computationally lightweight Multi-Layer Perceptron (MLP) policy network with two fully connected layers (64 neurons each),  ensuring efficient training and fast inference.
Performance is assessed across multiple scenarios, varying the number of GUs, UAVs, device capabilities, and grid size. Key environment and PPO parameters are defined in Table \ref{tab:combined_params}. 

\begin{table}[H]
\footnotesize
\centering
\caption{Environment and PPO Model Parameters}
\label{tab:combined_params}

\resizebox{8cm}{!}{
\begin{tabularx}{\columnwidth}{X X X X}
    \hline
    \textbf{Parameter} & \textbf{Value} & \textbf{Parameter} & \textbf{Value} \\
    \hline
    Grid size & 100×100 & $B_{ug}$ & 50 MHz \\
    O-RUs & 2 & $B_{ua}$ & 40 MHz \\
    UAVs & 3 & $B_{ag}$ & 100 MHz \\
    GUs & 10 & $E_{cp}$ & 4 W \\
    $\mathcal{T}$ & 10 & $E_{comm}$ & 7 W \\
    $M_a$ & 3 & $P_{ug}$ & 1 W \\
    $M_g$ & 3 & $P_{ua}$ & 2 W \\
    $W_g$ & [6, 12] & $P_{ag}$ & 4 W \\
    $\Gamma_u$ & [656, $1.7\times10^7$] & $D_{u,t}$ & [1, 10] MB \\
    $\mathcal{Q}_{u}$ & [1.8, 2.4] GHz & $\mathcal{P}_{g}$ & [3.5, 3.9] GHz \\
    $Z_u$ & [50, 250] J & $P_0$ & 30 W \\
    $P_1$ & 1.5 W & $c_0$ & 0.02 W/(m/s)$^3$ \\
    $\delta_t$ & 5 & $U_{\text{tip}}$ & 50 m/s \\
    $v_0$ & 30 m/s & Users mobility & Manhattan model \\
    \hline
\end{tabularx}
}

\resizebox{8cm}{!}{
\begin{tabularx}{\columnwidth}{X X X X}
    \hline
    \textbf{Parameter} & \textbf{Value} & \textbf{Parameter} & \textbf{Value} \\
    \hline
    Learning rate & $0.003$ & Number of epochs & $10$ \\
    Discount factor & $0.99$ & Batch size & $64$ \\
    Clip range & $0.2$ & Adjusting factor & $0.01$ \\
    \hline
\end{tabularx}
}
\end{table}

\begin{figure*}[t]
    \centering
    \begin{subfigure}[b]{0.24\textwidth}
        \centering
        \includegraphics[width=\textwidth]{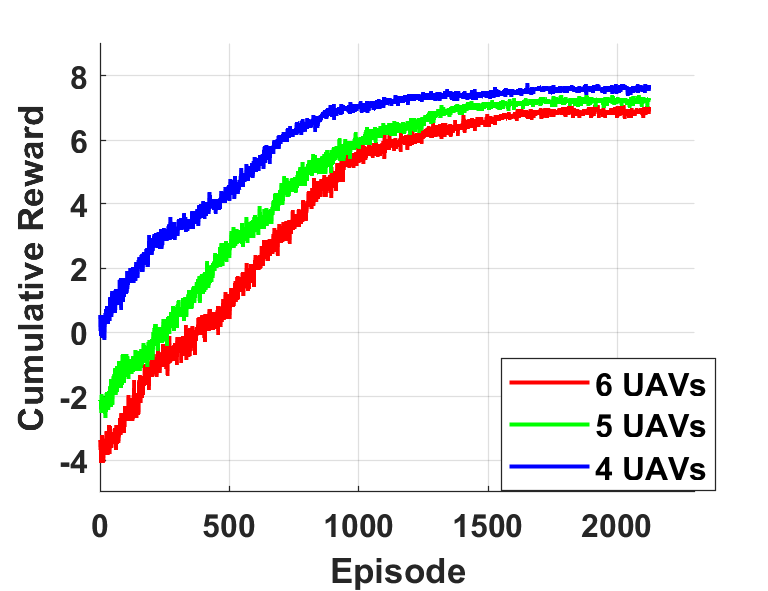}
        \captionsetup{justification=centering}
        \caption{}
        \label{fig:cum-reward}
    \end{subfigure}
    \hfill
    \begin{subfigure}[b]{0.24\textwidth}
        \centering
        \includegraphics[width=\textwidth]{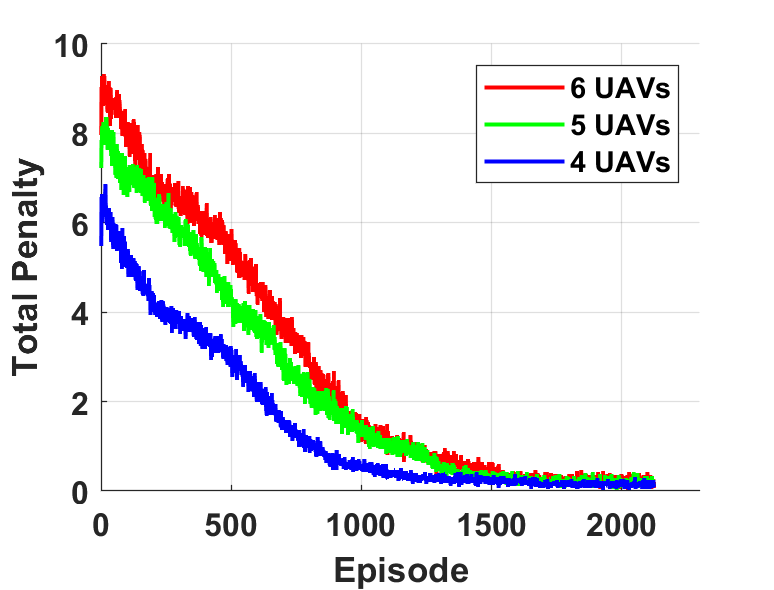}
        \captionsetup{justification=centering}
        \caption{}
        \label{fig:cum-loss}
    \end{subfigure}
    \hfill
    \begin{subfigure}[b]{0.24\textwidth}
        \centering
        \includegraphics[width=\textwidth]{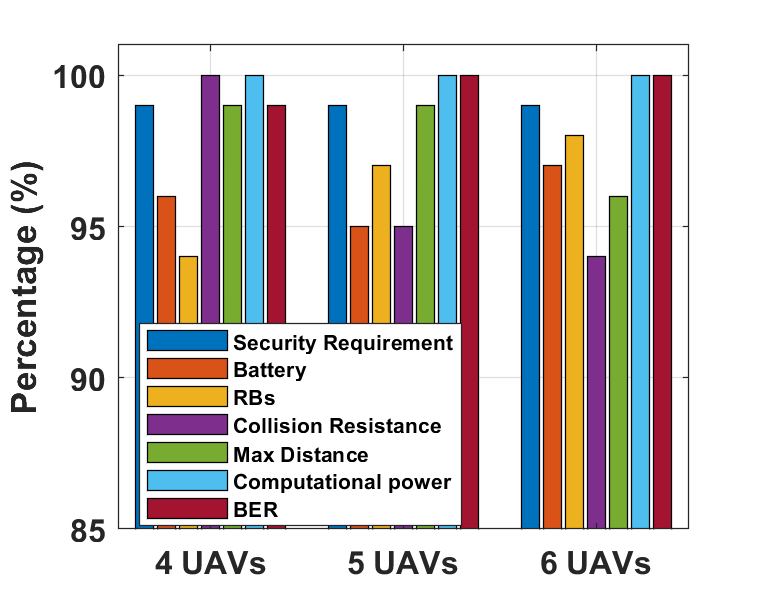}
        \captionsetup{justification=centering}
        \caption{}
        \label{fig:constraints_percentage}
    \end{subfigure}
    \hfill
    \begin{subfigure}[b]{0.24\textwidth}
        \centering
        \includegraphics[width=\textwidth]{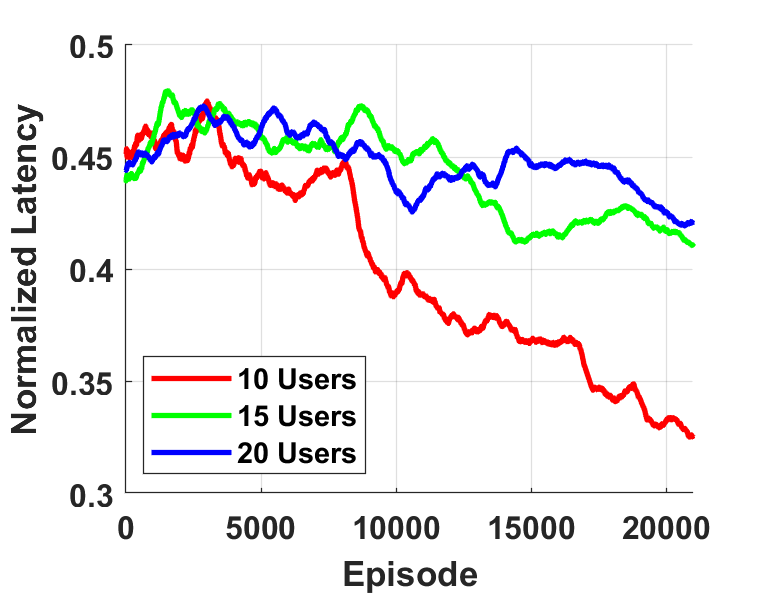}
        \captionsetup{justification=centering}
        \caption{}
        \label{fig:latency_numusers}
    \end{subfigure}

    \begin{subfigure}[b]{0.24\textwidth}
        \centering
        \includegraphics[width=\textwidth]{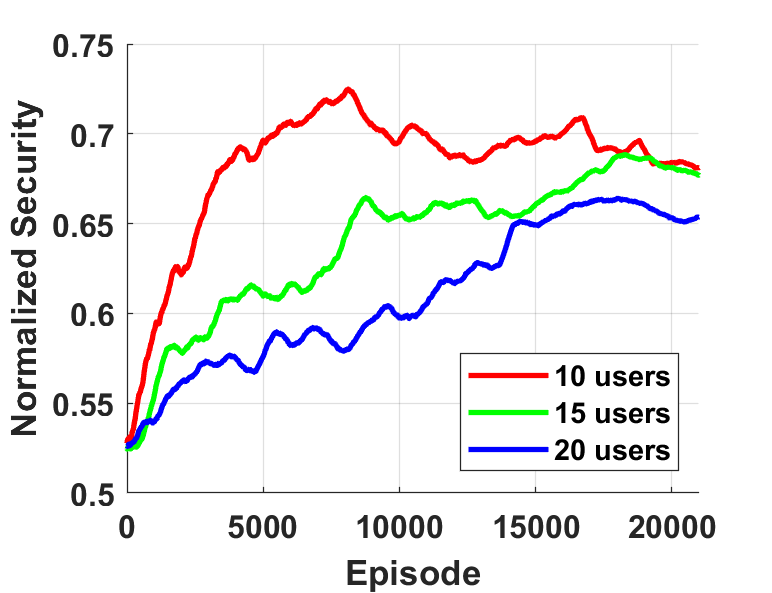}
        \captionsetup{justification=centering}
        \caption{}
        \label{fig:security_numusers}
    \end{subfigure}
    \hfill
    \begin{subfigure}[b]{0.24\textwidth}
        \centering
        \includegraphics[width=\textwidth]{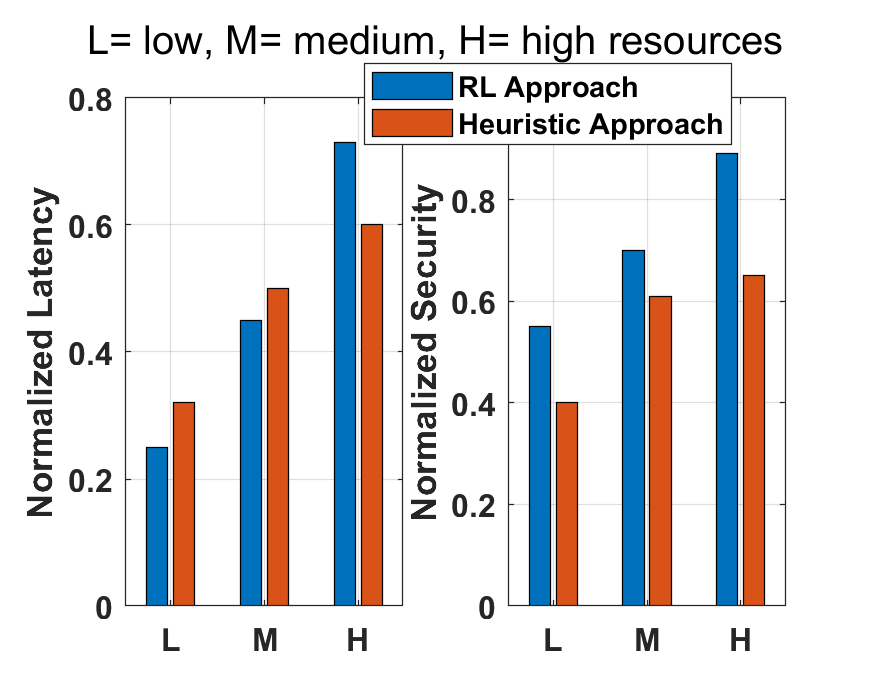}
        \captionsetup{justification=centering}
        \caption{}
        \label{fig:varying_capabilities}
    \end{subfigure}
    \hfill
    \begin{subfigure}[b]{0.24\textwidth}
        \centering
        \includegraphics[width=\textwidth]{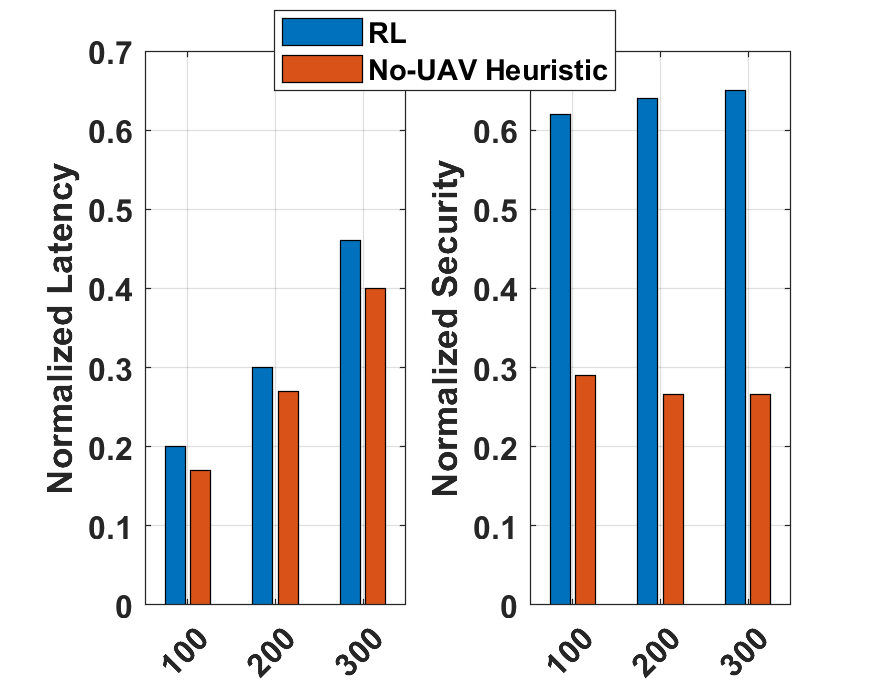}
        \captionsetup{justification=centering}
        \caption{}
        \label{fig:varying_grid_lat_Sec}
    \end{subfigure}
    \hfill
    \begin{subfigure}[b]{0.23\textwidth}
        \centering
        \includegraphics[width=\textwidth]{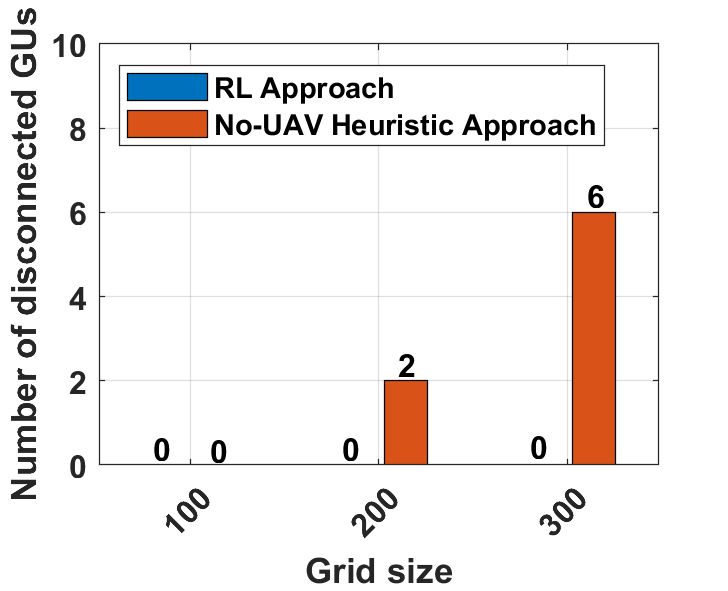}
        \captionsetup{justification=centering}
        \caption{}
        \label{fig:disconnected_devices}
    \end{subfigure}

    \caption{Performance of RL-based approach}
    \label{fig:all-results}
\end{figure*}

Figures (\ref{fig:cum-reward}) and \ref{fig:cum-loss} show the convergence of the proposed RL approach for different number of UAV relays (4, 5, and 6). Initially, random decisions lead to low rewards and constraint violations. As training progresses, Figures show a steady rise  in cumulative reward and decrease in cumulative penalty, indicating effective policy optimization and RL performance. Eventually, the reward curve plateaus, signaling convergence. 

Figure (\ref{fig:constraints_percentage}) shows the average percentage of constraints respected during the exploitation phase of the RL approach across different setups, calculated over 1,000 episodes. The results indicate that, on average, the RL model consistently adheres to constraints, with most values exceeding 95\%, highlighting effective enforcement of predefined limits. As the number of UAVs increases, the collision rate rises due to higher density, leading to more spatial conflicts. However, violations of RB constraints and BER decrease, as additional UAVs enhance resource allocation and signal quality. 

Figures (\ref{fig:latency_numusers}) and \ref{fig:security_numusers}) compare normalized latency and security across three setups with 10, 15, and 20 GUs and 3 UAV relays. As shown in Figure (\ref{fig:latency_numusers}), latency increases with the number of users, as UAVs become more loaded, reducing efficiency and increasing transmission time. In contrast, Figure (\ref{fig:security_numusers}) shows that security does not significantly improve with more users. Less users configuration achieved slightly higher security, likely due to a trade-off between latency and security. 

Figure (\ref{fig:varying_capabilities}) compares normalized latency and security between our RL approach and a heuristic baseline as device resources vary. The heuristic assigns each GU to the nearest available O-RU or UAV and places UAVs randomly while meeting minimum security requirements. The RL approach outperforms the heuristic in most cases. In low and medium-resource scenarios, RL achieves 10\% lower latency and 15\% higher security. In high-resource scenarios, RL shows slightly higher latency but over 20\% higher security, demonstrating its ability to prioritize security while balancing total latency. RL also optimizes UAV placement, reducing latency and mitigating delays by strategically positioning UAVs based on resource availability and traffic demand.

Figures (\ref{fig:varying_grid_lat_Sec}) and (\ref{fig:disconnected_devices}) compare the RL-based approach with the No-UAV heuristic across varying grid sizes. The RL approach outperforms the heuristic in security by utilizing UAV relays to maintain low BER and stable connections, while the No-UAV heuristic struggles as grid size increases. Latency rises for both, but in RL, this is due to higher security levels. Despite slightly higher latency, RL ensures full coverage. The No-UAV heuristic shows a gradual latency increase as devices disconnect, as shown in Figure (\ref{fig:disconnected_devices}), due to reliance on distant BSs, leading to higher BER and security failures. RL ensures stable communication and security by keeping all devices connected through UAV relays.

\vspace{-0.5em}
\section{Conclusion} \label{sec:conclusion}
This work addressed efficient resource management in UAV-based O-RAN systems, optimizing security, latency, and energy trade-offs through a novel RL-based framework for dynamic resource allocation. Designed for mission-critical scenarios like disaster response and SAR operations, the framework enhances security via encryption and resource optimization while minimizing latency and energy consumption. Simulations demonstrated significant improvements in security and efficiency, maintaining low latency even in resource-constrained environments.

\vspace{-1em}
\section{ACKNOWLEDGEMENT} \label{sec:acknowledgement}
This work was made possible by ARG grant - ARG01-0527-230356 from the Qatar National Research Fund (a member of Qatar Foundation). The findings achieved herein are solely the responsibility of the authors.
\vspace{-1em}
\bibliography{references}

\begin{thebibliography}{10}
\providecommand{\url}[1]{#1}
\csname url@samestyle\endcsname
\providecommand{\newblock}{\relax}
\providecommand{\bibinfo}[2]{#2}
\providecommand{\BIBentrySTDinterwordspacing}{\spaceskip=0pt\relax}
\providecommand{\BIBentryALTinterwordstretchfactor}{4}
\providecommand{\BIBentryALTinterwordspacing}{\spaceskip=\fontdimen2\font plus
\BIBentryALTinterwordstretchfactor\fontdimen3\font minus \fontdimen4\font\relax}
\providecommand{\BIBforeignlanguage}[2]{{%
\expandafter\ifx\csname l@#1\endcsname\relax
\typeout{** WARNING: IEEEtran.bst: No hyphenation pattern has been}%
\typeout{** loaded for the language `#1'. Using the pattern for}%
\typeout{** the default language instead.}%
\else
\language=\csname l@#1\endcsname
\fi
#2}}
\providecommand{\BIBdecl}{\relax}
\BIBdecl

\bibitem{uav_sar_survey}
\BIBentryALTinterwordspacing
M.~Lyu \emph{et~al.}, ``Unmanned aerial vehicles for search and rescue: A survey,'' \emph{Remote Sensing}, vol.~15, no.~13, 2023. [Online]. Available: \url{https://www.mdpi.com/2072-4292/15/13/3266}
\BIBentrySTDinterwordspacing

\bibitem{9839628}
A.~S. Abdalla \emph{et~al.}, ``Toward next generation open radio access networks: What o-ran can and cannot do!'' \emph{IEEE Network}, vol.~36, no.~6, pp. 206--213, 2022.

\bibitem{10584067}
H.~Li \emph{et~al.}, ``Energy-efficient deployment and resource allocation for o-ran-enabled uav-assisted communication,'' \emph{IEEE Transactions on Green Communications and Networking}, vol.~8, no.~3, pp. 1128--1140, 2024.

\bibitem{9522072}
D.-H. Tran \emph{et~al.}, ``Uav relay-assisted emergency communications in iot networks: Resource allocation and trajectory optimization,'' \emph{IEEE Transactions on Wireless Communications}, vol.~21, no.~3, pp. 1621--1637, 2022.

\bibitem{9712640}
Z.~Na \emph{et~al.}, ``Joint optimization of trajectory and resource allocation in secure uav relaying communications for internet of things,'' \emph{IEEE Internet of Things Journal}, vol.~9, no.~17, pp. 16\,284--16\,296, 2022.

\bibitem{10107729}
Z.~Lv \emph{et~al.}, ``Multi-agent reinforcement learning based uav swarm communications against jamming,'' \emph{IEEE Transactions on Wireless Communications}, vol.~22, no.~12, pp. 9063--9075, 2023.

\bibitem{Abughazzah2025}
Z.~Abughazzah \emph{et~al.}, ``Efficient resource management for secure and low-latency o-ran communication,'' in \emph{Wireless Communications and Networking Conference (WCNC)}.\hskip 1em plus 0.5em minus 0.4em\relax IEEE, 2025.

\bibitem{uav_energy}
\BIBentryALTinterwordspacing
Y.~Zeng \emph{et~al.}, ``Energy minimization for wireless communication with rotary-wing uav,'' 2018. [Online]. Available: \url{https://arxiv.org/abs/1804.02238}
\BIBentrySTDinterwordspacing

\bibitem{continuousDRL}
\BIBentryALTinterwordspacing
T.~P. Lillicrap \emph{et~al.}, ``Continuous control with deep reinforcement learning,'' 2019. [Online]. Available: \url{https://arxiv.org/abs/1509.02971}
\BIBentrySTDinterwordspacing

\bibitem{ppo_algorithms}
\BIBentryALTinterwordspacing
J.~Schulman \emph{et~al.}, ``Proximal policy optimization algorithms,'' 2017. [Online]. Available: \url{https://arxiv.org/abs/1707.06347}
\BIBentrySTDinterwordspacing

\end{thebibliography}
\bibliographystyle{IEEEtran}

\end{document}